\documentstyle[twocolumn,prl,aps,psfig]{revtex}

\begin{document}
\title{Thermal noise of a plano-convex mirror}
\author{A.\ Heidmann, P.F.\ Cohadon and M.\ Pinard\thanks{%
e-mail : heidmann, cohadon or pinard@spectro.jussieu.fr}}
\address{Laboratoire Kastler Brossel\thanks{%
Laboratoire de l'Universit\'{e} Pierre et Marie Curie et de l'Ecole Normale
Sup\'{e}rieure associ\'{e} au Centre National de la Recherche Scientifique},%
\\
Case 74, 4 place Jussieu, F75252\ Paris Cedex 05, France}
\date{September 1999}
\maketitle

\begin{abstract}
We study theoretically the internal thermal noise of a mirror coated on a
plano-convex substrate.\ The comparison with a cylindrical mirror of the
same mass shows that the effect on a light beam can be reduced by a factor
10, improving the sensitivity of high-precision optical experiments such
as gravitational-wave interferometers.
\end{abstract}

\bigskip

Thermal noise is a basic limit in many precision measurements\cite{Saulson90}%
. For example, the sensitivity in interferometric gravitational-wave
detectors\cite{Bradaschia90,Abramovici92} is limited by the Brownian motion
of the suspended mirrors.\ This thermally excited motion can be decomposed
into suspension and internal noises. The former refers to the motion of the
center of mass of the mirrors and it has been extensively studied both
theoretically\cite{Gonzalez94,Logan96} and experimentally\cite
{Logan92,Gillespie94,Gonzalez95,Braginsky96}.\ Internal noise is due to
thermally induced deformations of the mirror surface and constitutes the
main limitation of gravitational-wave detectors in the intermediate
frequency domain. It has thus been studied theoretically for mirrors
corresponding to the VIRGO\ and LIGO\ gravitational-wave observatories,
either by a decomposition of the motion over acoustic modes\cite
{Bondu95,Gillespie95} or by a direct approach based on the
fluctuation-dissipation theorem\cite{Levin98,Bondu98}. The normal-mode
expansion shows that the internal thermal noise may depend on the mirror
shape and on the spatial matching between light and acoustic modes.\ For
example, a bar-shaped cylindrical mirror is less noisy
than a gong-shaped one. A slight noise reduction has also been obtained
by moving the light beam off center. These works however deal with standard
cylindrical shapes although there is no evidence that such a geometry is the
best candidate to reach low thermal noise.

In this paper, we study an alternative geometry in which the mirror is
coated on the plane side of a plano-convex substrate. Such a plano-convex
geometry has been extensively studied for mechanical resonators with high
quality factors\cite{Wilson74}.\ Acoustic modes are confined near the
central axis of the resonator and their spatial structure can be described
by analytical expressions similar to gaussian optical modes of a Fabry-Perot
cavity\cite{Wilson74,Pinard99,Kogelnik66}. We show that this geometry leads
to a drastic reduction of the thermal noise.\ We first recall the effect of
mirror deformations on the light reflected by the mirror.\ We then determine
the mirror motion induced by thermal excitation and show that the effect on
light at low frequency can be expressed in terms of an effective
susceptibility which takes into account every acoustic mode and its
coupling to the light.\ We compare the effect of thermal noise to the one
obtained with a cylindrical mirror of the same mass.\ We finally determine the
noise reduction obtained by moving the light beam off center and we show
that a global noise reduction of 10\ can be reached.

\section{Effect of thermal noise on light}

\begin{figure}
\centerline{\psfig{figure=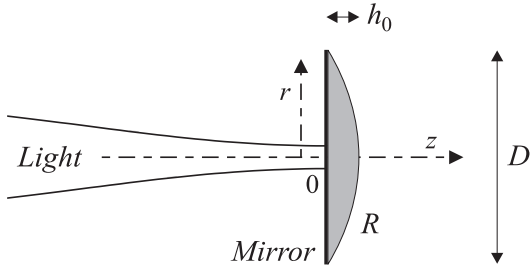,width=7cm}}
\vspace{2mm}
\caption{The mirror is coated on the plane side of a plano-convex sustrate of radius
$R$, thickness $h_{0}$ and diameter $D$.\ The effect of the thermally excited motion
corresponds to a global phase-shift for the light beam.}
\label{Fig_Model}
\end{figure}

In a gravitational-wave interferometer, the thermal noise of the mirrors has
a similar effect as a gravitational wave since both change the optical path
followed by the light in the interferometer arms.\ At every radial point $r$
of the mirror surface, the field experiences a local phase-shift
proportional to the longitudinal displacement $u\left( r,z=0,t\right) $ of
the surface (the origin of the cylindrical coordinates is taken at the
center of the mirror surface, see figure \ref{Fig_Model}). This leads to a
global phase-shift for the reflected field which is actually related to the
mirror displacement averaged over the beam profile\cite{Gillespie95,Pinard99}%
.\ The phase shift between the two interferometer arms thus contains
information about the mirror displacement and the variable read out by this
procedure corresponds to the averaged displacement 
\begin{equation}
\widehat{u}\left( t\right) =\left\langle u\left( r,z=0,t\right) ,v\left(
r\right) \right\rangle ,  \label{Eq_MeanU}
\end{equation}
where the brackets stand for the overlap integral in the mirror plane ($z=0$%
), 
\begin{equation}
\left\langle f\left( r\right) ,g\left( r\right) \right\rangle
=\int_{z=0}d^{2}rf\left( r\right) g\left( r\right) ,  \label{Eq_Overlap}
\end{equation}
and $v\left( r\right) $ is the intensity profile of the light beam in the
mirror plane.\ Assuming that the beam is in a TEM$_{00}$ Gaussian mode, this
profile is related to the beam waist $w_{0}$ by 
\begin{equation}
v\left( r\right) =\frac{2}{\pi w_{0}^{2}}e^{-2r^{2}/w_{0}^{2}}.  \label{Eq_v}
\end{equation}

Any displacement can be decomposed onto the acoustic modes of the mirror.\
Noting $\left\{ u_{n}\left( r,z\right) \right\} $ a basis of the internal
acoustic modes, the displacement $u\left( r,z,t\right) $ can be expressed as
a linear combination of these modes 
\begin{equation}
u\left( r,z,t\right) =\sum_{n}a_{n}\left( t\right) u_{n}\left( r,z\right) ,
\label{Eq_Decomp}
\end{equation}
where $a_{n}\left( t\right) $ is the time-dependent amplitude of mode $n$.
Each acoustic mode corresponds to a harmonic oscillator characterized by a
Lorentzian mechanical susceptibility 
\begin{equation}
\chi _{n}\left[ \Omega \right] =\frac{1}{M_{n}\left( \Omega _{n}^{2}-\Omega
^{2}-i\Omega _{n}^{2}\Phi \left[ \Omega \right] \right) },  \label{Eq_ChiN}
\end{equation}
where $M_{n}$ is the effective mass of mode $n$, $\Omega _{n}$ is its
resonance frequency and $\Phi \left[ \Omega \right] $ is the loss angle
assumed to be the same for all modes. The mirror motion can be described by
the Fourier transforms $a_{n}\left[ \Omega \right] $ of every amplitude
coefficient.\ Assuming that the mirror is in thermal equilibrium at
temperature $T$, one gets 
\begin{equation}
a_{n}\left[ \Omega \right] =\chi _{n}\left[ \Omega \right] F_{T,n}\left[
\Omega \right] ,  \label{Eq_an}
\end{equation}
where $F_{T,n}$ is a Langevin force describing the coupling of the mode $n$
with the thermal bath.

One can now determine the averaged displacement $\widehat{u}$ from eqs.\ (%
\ref{Eq_MeanU}), (\ref{Eq_Decomp}) and (\ref{Eq_an}).\ One gets\cite{Pinard99} 
\begin{equation}
\widehat{u}\left[ \Omega \right] =\chi _{eff}\left[ \Omega \right] F_{T}%
\left[ \Omega \right] ,  \label{Eq_uChi}
\end{equation}
where $\chi _{eff}$ appears as an effective susceptibility taking into
account all the acoustic modes and their spatial overlap with the light
beam, 
\begin{equation}
\chi _{eff}\left[ \Omega \right] =\sum_{n}\left\langle u_{n}\left(
r,z=0\right) ,v\left( r\right) \right\rangle ^{2}\chi _{n}\left[ \Omega %
\right] .  \label{Eq_ChiEff}
\end{equation}
This effective susceptibility is then the sum over all modes of the
susceptibilities $\chi_{n}$ weighted by the overlap with light. The
force $F_{T}$ in eq. (\ref{Eq_uChi}) is an effective Langevin force related
to the forces $F_{T,n}$ of each acoustic mode.\ One finds\cite{Pinard99} that
the noise spectrum $S_{T}\left[ \Omega \right] $ of the force $F_{T}$ is
related to $\chi _{eff}$ by the fluctuation-dissipation theorem 
\begin{equation}
S_{T}\left[ \Omega \right] =-\frac{2k_{B}T}{\Omega }%
\mathop{\rm Im}%
\left( \frac{1}{\chi _{eff}\left[ \Omega \right] }\right) .  \label{Eq_ST}
\end{equation}
This relation means that the mirror is in thermodynamic equilibrium at
temperature $T$.

In a gravitational-wave interferometer, the frequency of a gravitational
wave is usually much smaller than the mechanical resonance frequencies of
internal acoustic modes of the mirrors. We are thus interested in the noise
spectrum $S_{\widehat{u}}$ of the averaged displacement $\widehat{u}$ at low
frequency compared to the resonance frequencies $\Omega _{n}$. From eqs. (%
\ref{Eq_uChi}) to (\ref{Eq_ST}), the background thermal noise can be
approximated in this frequency domain to 
\begin{eqnarray}
S_{\widehat{u}}\left[ \Omega \approx 0\right] &=&\frac{2k_{B}T}{\Omega }%
\mathop{\rm Im}%
\left( \chi _{eff}\left[ \Omega \right] \right)  \nonumber \\
&\approx &2k_{B}T\frac{\Phi \left[ \Omega \right] }{\Omega }\chi _{eff}\left[
0\right] .  \label{Eq_Su}
\end{eqnarray}
The effect of thermal noise on light is thus proportional to the effective
susceptibility at zero frequency which is given by 
\begin{equation}
\chi _{eff}\left[ 0\right] =\sum_{n}\frac{\left\langle u_{n}\left(
r,z=0\right) ,v\left( r\right) \right\rangle ^{2}}{M_{n}\Omega _{n}^{2}}.
\label{Eq_ChiEff0}
\end{equation}
In the next section we determine this susceptibility for a plano-convex
mirror and we compare the values obtained to the ones of a cylindrical
mirror.

\section{Plano-convex mirror}

We consider that the mirror is coated on the plane side of a segment of
sphere of radius $R$ and of thickness $h_{0}$ (see figure \ref{Fig_Model}).
For simplicity we assume that the mirror has a sharp edge on its
circumference so that its mass $M$ and its diameter $D$ are related to $R$
and $h_{0}$ by 
\begin{eqnarray}
M &=&\pi \rho h_{0}^{2}\left( R-h_{0}/3\right) ,  \label{Eq_M} \\
D &=&2\sqrt{h_{0}\left( 2R-h_{0}\right) },  \label{Eq_D}
\end{eqnarray}
where $\rho $ is the density of the substrate (2200 kg/m$^{3}$ for silica).
The total mass $M$ of the mirror is an important parameter for the
suspension thermal noise\cite{Gonzalez94,Logan96}. We thus choose a mass of
the same order as the one of the mirrors in gravitational-wave
interferometers.\ We will see however that the internal thermal noise is
quite insensitive to the mirror mass so that we set the mass $M$ to 20 kg.
All the geometrical parameters of the mirrors (curvature radius $R$,
diameter $D$) can then be expressed in terms of the thickness $h_{0}$.

If the thickness is much smaller than the curvature radius, the acoustic
propagation equation can be solved using a paraxial approximation and one
gets analytical expressions for the acoustic modes corresponding to Gaussian
modes\cite{Wilson74}. Each compression mode is defined by three integers $n$%
, $p$, $l$, corresponding to longitudinal, radial and angular indexes,
respectively.\ The longitudinal displacement $u_{n,p,l}\left( r,z\right) $
at radial coordinate $r$ and axial coordinate $z$ is given by\cite
{Wilson74,Pinard99} 
\begin{equation}
u_{n,p,l}\left( r,z\right) =e^{-r^{2}/w_{n}^{2}}L_{p}^{l}\left(
2r^{2}/w_{n}^{2}\right) \cos \left( \frac{n\pi }{h\left( r\right) }z\right) .
\label{Eq_unpl}
\end{equation}
$u_{n,p,l}$ is composed of a transverse Gaussian structure, a transverse
Laguerre polynomial $L_{p}^{l}$ and a cosine in the propagation direction ($%
h\left( r\right) $ is the mirror thickness at radial position $r$, equal to $%
h_{0}$ for $r=0$). The acoustic waist $w_{n}$ and the eigenfrequency $\Omega
_{n,p,l}$ are given by 
\begin{eqnarray}
w_{n}^{2} &=&\frac{2h_{0}}{n\pi }\sqrt{Rh_{0}},  \label{Eq_wn} \\
\Omega _{n,p,l}^{2} &=&\Omega _{M}^{2}\left[ n^{2}+\frac{2}{\pi }\sqrt{\frac{%
h_{0}}{R}}n\left( 2p+l+1\right) \right] ,  \label{Eq_Onpl}
\end{eqnarray}
where $\Omega _{M}=\pi c_{l}/h_{0}$ is the fundamental longitudinal
frequency, $c_{l}$ being the longitudinal sound velocity (5960 m/s for
silica).

From these equations one can derive an analytical expression for the
effective susceptibility $\chi _{eff}\left[ 0\right] $ as an infinite sum
over all modes $\left\{ n,p,l\right\} $ (eq.\ \ref{Eq_ChiEff0}). In the case
where the light beam is centered on the mirror, only modes that have a
cylindrical symmetry will contribute.\ In particular the sum over $l$ is
limited to $l=0$. The effective mass of each acoustic mode and the spatial
overlap with light are then given by 
\begin{eqnarray}
M_{n} &=&\frac{\pi }{4}\rho h_{0}w_{n}^{2},  \label{Eq_Mn} \\
\left\langle u_{n,p,0}\left( r,z=0\right) ,v\left( r\right) \right\rangle &=&%
\frac{2w_{n}^{2}}{2w_{n}^{2}+w_{0}^{2}}\left( \frac{2w_{n}^{2}-w_{0}^{2}}{%
2w_{n}^{2}+w_{0}^{2}}\right) ^{p}.  \label{Eq_Ovl}
\end{eqnarray}

\begin{figure}
\centerline{\psfig{figure=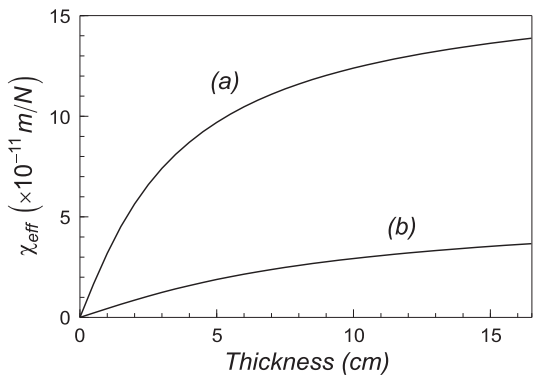,width=7cm}}
\vspace{2mm}
\caption{Variation of the effective susceptibility at zero frequency
$\chi _{eff}\left[ 0\right] $ as a function of the thickness $h_{0}$ of the mirror.\ Curves
(a) and (b) correspond to an optical waist $w_{0}$ of 2 cm and 5.5 cm, respectively.}
\label{Fig_Thick}
\end{figure}

We have numerically computed the effective susceptibility for different
thicknesses.\ Figure \ref{Fig_Thick} shows the result obtained by computing 30
different values of the thickness.\ For each value, the curvature radius $R$
and the diameter $D$ of the mirror are determined according to eqs. (\ref
{Eq_M}) and (\ref{Eq_D}). For example one gets $R=61$ cm and $D=57$ cm for a
thickness $h_{0}$ of 7 cm. The two curves in figure \ref{Fig_Thick} are obtained with
different optical waists $w_{0}$ (2 cm for curve {\it a} and 5.5 cm for
curve {\it b}). These waists correspond to the beam waists on the front and
end mirrors of the VIRGO\ interferometer\cite{Bondu95}. One observes a
decrease of the thermal noise for a thinner mirror.\ This is partly due to
the fact that the mechanical resonance frequencies are increased.\ It would
be however difficult to use a very thin mirror since its diameter would
become very large ($D$ evolves as $h_{0}^{-1/2}$ for small $h_{0}$).

If we consider a reasonable thickness of 7 cm, we obtain an effective
susceptibility $\chi _{eff}\left[ 0\right] $ equal to $11\times 10^{-11}$
m/N for an optical waist of 2 cm, and $2.4\times 10^{-11}$ m/N for $%
w_{0}=5.5 $ cm. These results can be compared to the values obtained for
cylindrical mirrors, that is $46\times 10^{-11}$ m/N ($w_{0}=2$ cm) and $%
11\times 10^{-11}$ m/N ($w_{0}=5.5$ cm)\cite{Bondu95}. The internal thermal
noise of a plano-convex mirror is thus significantly smaller, by at least a
factor 4. If the constraint on the diameter can be relaxed, even larger
noise reduction may be obtained.

\begin{figure}
\centerline{\psfig{figure=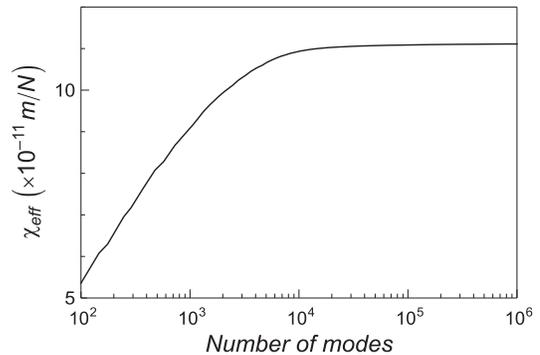,width=7cm}}
\vspace{2mm}
\caption{Convergence of the effective susceptibility as a function of the number of
computed modes.}
\label{Fig_Precis}
\end{figure}

We have checked the validity of the numerical calculation by plotting the
effective susceptibility as a function of the number of computed modes, for
example in the case of a thickness of 7 cm and an optical waist of 2 cm
(figure \ref{Fig_Precis}).\ This curve shows that results become valid as soon
as the number of computed modes is larger than 10$^{4}$. Since the numerical
calculation only deals with simple analytical expressions, it can easily be
processed with a very large number of modes, such as 10$^{6}$.

\section{Relation with the optical mass}

\begin{figure}
\centerline{\psfig{figure=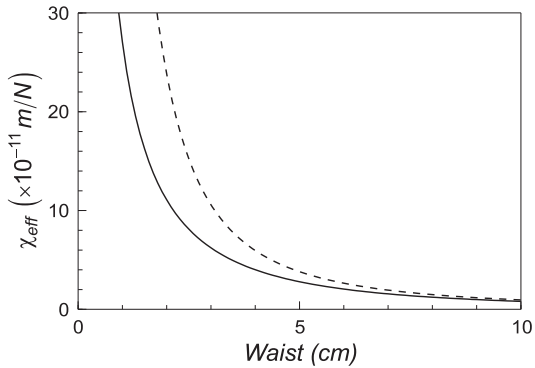,width=7cm}}
\vspace{2mm}
\caption{Variation of the effective susceptibility as a function of the optical waist
$w_{0}$ for a 7-cm thick mirror. The solid curve is the computed result and the dashed
curve corresponds to an approximation for which the effective mass of the mirror is
replaced by the optical mass.}
\label{Fig_Waist}
\end{figure}

Figure \ref{Fig_Waist} shows the variation of the effective susceptibility
as a function of the optical waist. The thermal noise is reduced for a wider
waist.\ The mirror displacement is actually averaged over the beam waist.\
Since the maximum displacement is at the center of the mirror, one gets less
noise for a wide waist.

It is possible to derive a simple approximation of the thermal noise in the
case of a thickness $h_{0}$ much smaller than the curvature radius $R$. We
can then assume that the transverse acoustic modes are degenerate and we can
replace the resonance frequencies $\Omega _{n,p,0}$ by the value for $p=0$
(eq.\ \ref{Eq_Onpl}).\ The sum over $p$ in the effective susceptibility
(eq.\ \ref{Eq_ChiEff0}) is then a geometric sum and one gets a simple
estimate of the susceptibility $\chi _{eff}\left[ 0\right] $ in terms of an
optical mass $M_{opt}$ given by\cite{Pinard99} 
\begin{eqnarray}
\chi _{eff}\left[ 0\right] &\approx &1/M_{opt}\Omega _{M}^{2},
\label{Eq_ChiEffApprox} \\
M_{opt} &=&\frac{12}{\pi ^{2}}\left( \frac{\pi }{4}\rho
h_{0}w_{0}^{2}\right) .  \label{Eq_Mopt}
\end{eqnarray}
The effective susceptibility is thus equivalent to the one of a single
harmonic oscillator of resonance frequency $\Omega _{M}$ and of mass $%
M_{opt} $ corresponding to the mass of the part of the mirror illuminated by
the light beam (term in brackets in eq.\ \ref{Eq_Mopt}). The resulting
susceptibility is shown as a dashed curve in figure \ref{Fig_Waist} and
appears to be a good approximation of the computed susceptibility. It is
actually an overestimation of the effective susceptibility since the
resonance frequencies $\Omega _{n,p,0}$ are always larger than $\Omega
_{n,0,0}$.

This result shows that the internal thermal noise depends only on a few
parameters, mainly the optical waist $w_{0}$ and the thickness $h_{0}$. The
effective susceptibility evolves as $h_{0}/w_{0}^{2}$ and one obtains a lower
noise for a thin mirror and a wide optical waist, as shown in figures \ref
{Fig_Thick} and \ref{Fig_Waist}. We have ckecked that the thermal noise is
quite insensitive to other parameters such as the curvature radius $R$ or
the mass $M$ of the mirror.\ We have computed the effective susceptibility
for different masses by varying the curvature radius $R$ while the thickness
$h_{0}$ is kept constant. The relative variation of the thermal noise is less
than 5\% for masses between 5 kg and 50 kg.

Note that this relation with the optical mass seems to be a specific
behavior of the plano-convex geometry.\ One can define an effective mass for
a cylindrical mirror by a relation similar to (\ref{Eq_ChiEffApprox}) but
this mass is no longer related to the optical mass.\ From the point of view
of thermal noise, the comparison between cylindrical and plano-convex
geometries shows that the effective mass of a cylindrical mirror is usually
larger than the optical mass, whereas the fundamental resonance frequency $%
\Omega _{M}$ is approximately 10\ times larger for the plano-convex geometry
(typically 40 kHz instead of 4 kHz).\ This drastic increase of the
fundamental frequency explains why the plano-convex mirror is less noisy.

\section{Misalignment of the light beam}

We now study the effect of a misalignment of the light beam. Since the
maximum thermal displacement is at the center of the mirror, moving the light
beam off center must improve the noise.\ However, the number of modes ($l\neq 0$%
) which contributes to the thermal noise rapidly increases with the
misalignment. As a consequence there is only a slight effect for
cylindrical mirrors\cite{Bondu95}. In the case of a plano-convex mirror,
acoustic modes are more confined around the center of the mirror and one
expects a larger effect of misalignment.

\begin{figure}
\centerline{\psfig{figure=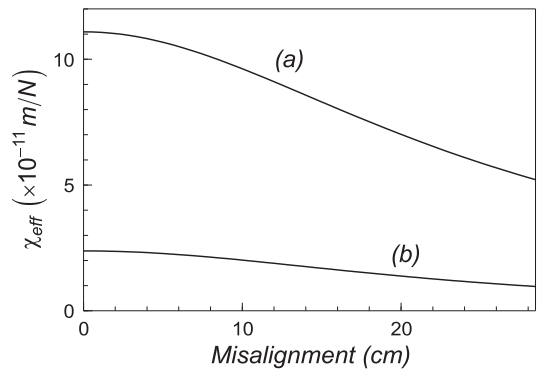,width=7cm}}
\vspace{2mm}
\caption{Variation of the effective susceptibility as a function of the misalignment of the
light beam. Curves (a) and (b) correspond to an optical waist $w_{0}$ of 2 cm and
5.5 cm, respectively.}
\label{Fig_Align}
\end{figure}

To efficiently compute the effective susceptibility, we replace the Laguerre
polynomial in eq.\ (\ref{Eq_unpl}) by two Hermite polynomials and we use the
recurrence relations between them in the numerical calculation.\ The
resulting susceptibility is shown in figure \ref{Fig_Align}, for a 7-cm
thick mirror and for an optical waist of 2 cm (curve {\it a}) and of 5.5\ cm
(curve {\it b}). The reduction of thermal noise with misalignment is clearly
visible on those curves since a reduction factor on the order of 2 can be
reached.

\section{Conclusion}

We have studied the internal thermal noise at low frequency of a
plano-convex mirror. We have shown that it can be related to the
susceptibility of an equivalent pendulum of resonance frequency equal to the
fundamental resonance frequency of the mirror and of mass equal to the
optical mass.\ As a consequence the thermal noise mainly depends on the
thickness of the mirror and on the waist of the light beam. We have found that
a thin plano-convex mirror has a lower noise than a standard cylindrical
mirror of same mass. For a 7-cm thickness, the thermal noise is approximately
10\ times lower.\ Using such plano-convex mirrors in gravitational-wave
interferometers would thus increase the sensitivity of the interferometer in
the intermediate frequency domain.

The results derived in this paper are based on a paraxial approximation which
allows to describe the acoustic modes of the mirror as Gaussian modes. This
approximation is valid if the diameter and the curvature radius of
the mirror are large as compared to its thickness and to the optical
waist. Otherwise, high-order transverse modes may become sensitive to
edge effects and they may be not adequately described as Gaussian modes. These
modes however do not significantly contribute to the thermal noise since the
overlap with light becomes small for high-order modes. Anyway, considering the
large noise reduction that would be obtained by a simple change of the mirror
shape, we believe that it would be of great interest to experimentally compare
the background thermal noises of cylindrical and plano-convex mirrors, for example
by using a high-sensitivity displacement sensor such as a high-finesse
cavity\cite{Hadjar99}.

Finally, let us note that we have studied a geometry with a sharp edge on the
circumference.\ Since the acoustic modes are confined near the center of the
mirror, it is possible to truncate the borders of the mirror in order to
have an edge with a finite size. This would not degrade the thermal
characteristics of the mirror, but it will change its diameter and its mass.
A compromise must be found in practice between the thickness which
determines the amplitude of internal noise, the mass which is important for
suspension thermal noise, and the diameter of the mirror.

\end{document}